# Graphene resistivity in diffusive limit due to scalar and vector potential electron-phonon scattering

**M Obaidurrahman and S S Z Ashraf[*]**

*Department of Physics, Faculty of Science, Aligarh Muslim University Aligarh- 202002, UttarPradesh, India*

Email; ssz_ashraf@rediffmail.com

## Abstract

We calculate the contribution of unscreened and screened scalar (DP) and vector potential(VP) electron-acoustic phonon coupling to resistivity in disordered graphene through Keldysh Green's function method within the diffusive limit, $k_F l >> 1$. We obtain analytical results in the asymptotic limits of clean and strong impurities for both DP and VP coupling, which is in observed to be in good agreement with the resistivity behavior in these limits. We find that the complete numerical results approach the analytical result in the extreme limits, but give different temperature dependencies in between these limits. The graphene resistivity has been investigated as functions of temperature, mean free path and carrier density. We also evaluated the screened behavior in the Thomas-Fermi and Random phase approximation dielectric function and obtained the temperature power exponents. We find that in the absence of screening when the electronic disorder ($T<T_{dr}$), the two coupling mechanisms are affected differently and the relaxation rate associated with the VP coupling is suppressed by disorder as compared to the DP coupling, and the DP coupling is enhanced by disorder.

Keywords; Graphene, Disorder, Deformation Potential, Vector Potential, Resistivity, Keyldysh Method

## 1. Introduction

Ever since its very emergence on the materials' landscape, the impact of the disorder on both thermal and electric transport properties in graphene has been an important aspect of research, as it is directly linked to the manifestation of many novel phenomena and device applications. In graphene, the intrinsic and extrinsic disorders such as lattice defects, impurities, charge traps, resonant states, ripples, adatoms/molecules, Stone-Thrower-Wales defects etcetera can exist and influence several observables particularly the transport properties [1-5]. Among these varieties of disorder, the one or many of which is/are going to affect the properties of the sample at large will depend on synthesis technology. Physically, this is because of the fact that in graphene the novel massless Dirac charge carriers in collusion with disorder entails the observation of extremely rich and unique transport phenomena such as Klein tunneling, the famous minimum conductivity at the Dirac point, anomalous quantum Hall effect, weak anti localization, disorder limited thermal cooling, etcetera [1-6]. Graphene turns out to be the only system in which disorder assisted cooling dominates from very low to the room temperature [6]. The disorder driven topological properties in graphene stands out as a unique area in graphene with its own rich repercussions [7]. The effects of the disorder are not always detrimental as has been amply portrayed in the case of graphene where the presence of disorder combined with a weak electron-phonon (e-ph) coupling could be utilized in a multitude of technological applications like the engineering of mobility gaps, calorimetry, bolometry, infrared, and THz detectors, Biosensors, etcetera [8-12].

The fact that the e-ph coupling is weak enough in graphene makes the conduction electrons and the lattice stay at significantly different temperatures for a longer span of time [13]. The e-ph coupling gives a measure of energy that can be transferred from hot electrons to phonons, and in thermal metrology it enables the measurement of phonon temperature if the in equilibrium persists for a sufficient period of time. A recent theory of the e-ph coupling has predicted a large modification of the e-ph coupling due to the effects of disorder and scattering of the electrons [6]. The effects of the disorder are varied and depend on the nature of the disorder and method of sample preparation in general, and so is the case in graphene [6, 13]. For instance the heat flow from the hot electrons to phonons in general is described by the power transfer law having the form $P = \Upsilon\Omega(T_e^n - T_{ph}^n)$ where $\Upsilon$ is the e-ph coupling constant, $\Omega$ the area of graphene, $T_e$ the electron temperature, $T_{ph}$ the phonon temperature, and $n$ the temperature power exponent. The e-ph coupling in graphene is dependent upon temperature, carrier density, substrate, screening, and disorder [13-17]. The coupling constant strength $\Upsilon$ and the value of exponent $n$ depend on disorder. The effects of disorder are indistinct as the power law exponent $n$ can either be enhanced by disorder or be suppressed depending upon on the nature of the disorder and the electronic system. Theoretical and experimental works have established $n$=4 in clean graphene, but in disordered graphene conflicting reports indicating either way transition in power exponent of $n = 3 \leftrightarrows 4$ have been reported [18-19]. In a weakly disordered graphene that is $k_F l \gg 1$, where $k_F$ is the Fermi momentum and $l$ is the mean free path, a possibility of $n$ ranging from 3 to 6 has also been predicted [20]. With no established consensus in either theory or experiments on transport properties, disordered graphene is still a subject of immense debate and scientific exploration.

Similarly, the resistivity of disordered materials shows surprising temperature dependence at low temperatures. Though few in number, conflicting reports on resistivity/conductivity of graphene with disorder also exist. Experimental data on these effects have been very limited because of the high detection sensitivity needed for such measurements. In the clean limit with DP coupling, many theoretical works have reported the same power dependency for resistivity/e-ph relaxation rate, i.e.( $T^4$ ) and power loss rate [21-24], which has been confirmed in the low-temperature limit in various experimental works also [25-26]. An alternate mechanism of super collisions in the presence of disorder has been invoked to overcome the limitations of momentum-conserving e-ph normal scattering for the explanation of heat flow in disordered graphene [6]. A recent study to estimate the e-ph mediated heat flow in disordered graphene using the Keyldysh method has been published [20]. Though the temperature dependency of resistivity in the diffusive limit for disordered graphene (with the short and long-range electronic disorder) in the Boltzmann transport regime have been reported in a couple of studies [22-23] but the behavior of resistivity of disordered graphene with deformation or scalar potential coupling (DP) and vector or gauge potential coupling (VP) coupling has not been reported so far.

At low temperatures, the acoustic e-ph scattering is the dominant source of scattering as the optical phonons have very high energy and their coupling to electrons is very weak and therefore insignificant for carrier transport at low temperatures [16]. The transport properties such as energy loss rate, thermo power, mobility, resistivity, etc. are in general accounted by the DP but in graphene or in any 2D dimensional crystal a VP coupling also independently arises due to ripples. The 2D sheets of crystals are known to develop static shape fluctuations due to the unavoidable thermodynamic instability with respect to both crumpling and bending. The DP coupling stems from the local dilation of the lattice and the ripples originate from a pure shear deformation of the lattice. The ripples can be alternatively seen as a static random gauge potential that corresponds to a change in hopping matrix elements accompanying the shear deformation, and appearing in the Dirac Hamiltonian as an external gauge field term [6]. In the presence of the static potential, quantum interference of numerous scattering processes drastically changes the effective e-ph interaction. In graphene, the chiral nature of Dirac carriers further modifies the role of electronic diffusion, e-ph DP, and VP coupling [20].

Several theoretical approaches like Boltzmann Transport theory [23], Hydrodynamic approach [27], Green's function-based Kubo method [20], apart from simulation methods [28], are used to describe electron transport properties of the disordered graphene, but the simplest approach is that of Boltzmann transport formalism (BTF) in which the scattering is treated within the Fermi Golden rule. Though the BTF is particularly effective in the pure system in thermal equilibrium and for scattering from charge impurities, it is of limited applicability for other types of the disorder [29]. Now it is well known that disorder scattering induces quantum interference between self-crossing paths in mesoscopic samples [30]. Therefore in disordered systems, the modifications arising out of quantum corrections that complicate the calculations need to be addressed. The theoretical approach in accounting corrections arising from this interference is achieved through the Keyldysh method based on non-equilibrium Green's function method [31]. The Keyldysh non-equilibrium Green's function technique is versatile enough to capture the vertex corrections of the e-ph interaction influenced by the disorder. Also, the DP e-ph coupling is expected to be screened whereas the VP coupling is suggested to be unscreened as it induces no net electronic charge. This is because the effective VP generated by a phonon field has the opposite sign in the two graphene valleys [32-33]. Therefore in this work, we investigate and calculate both the screened and unscreened e-ph interaction dependent relaxation rate and resistivity in single-layer graphene in the weakly disordered regime, $k_F l >> 1$ in the Keldysh formalism.

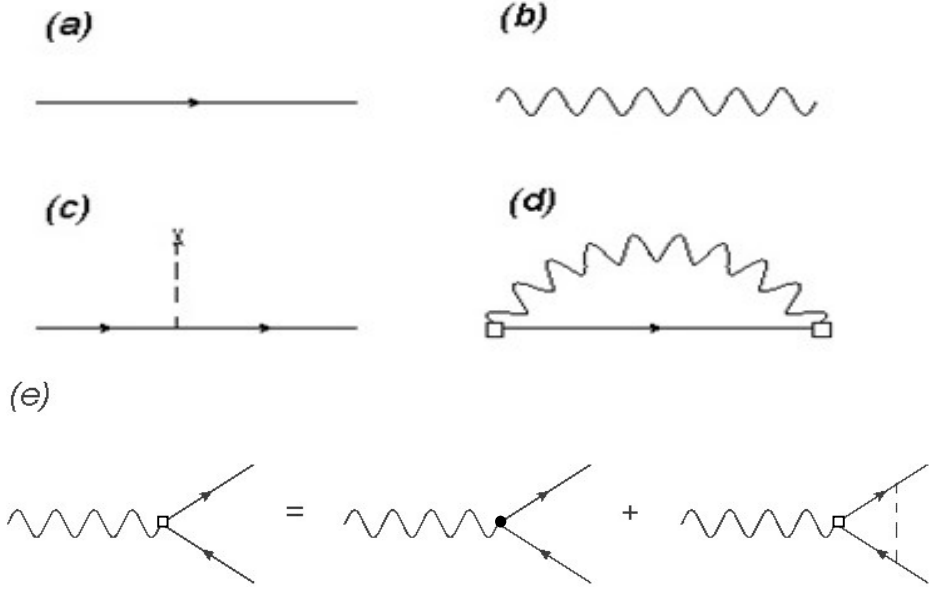

**Figure 1.** Feynman diagrams (a) Electron propagator, (b) Phonon propagator,(c) Electron-impurity propagator for a single scattering event- the dashed lines with cross represent impurity potential, (d) Electron-phonon self-energy propagator-the square block represents diffusion dressed vertex, (e) Vertex correction of the electron-phonon interaction by impurity scattering. An impurity average denoted by the dashed line while dressed vertex and bare vertex represented by the square block and dot respectively. The figures (d) & (e) are adapted from Ref. [20].

## 2. Formalism

The honeycomb graphene lattice comprises two equivalent triangular sub lattices contributing two atom sites per unit cell. The effective disordered graphene Hamiltonian [34] in presence of e-ph interaction is

$$H = H_e + H_{ph} + H_{dis} + H_{e-ph} \quad (1)$$

Where the non-interacting tight binding Hamiltonian $H_e = \sum_{\mathbf{k}} \psi_{\mathbf{k}}^{\dagger} (-i v_f \sigma_i . \nabla_i) \psi_{\mathbf{k}}$ corresponds to the low energy $\pi$ electrons near the K points, $v_f$ is the Fermi velocity, $\psi_{\mathbf{k}}, \psi_{\mathbf{k}}^{\dagger}$ are the two component spinors from the two sub lattice space and $\sigma_i (i=x, y)$ are the Pauli matrices operating in the same space. This non-interacting Hamiltonian $H_e$ yields an effective linear energy dispersion relation, $\epsilon(\mathbf{k}) = \pm \hbar v_f |\mathbf{k}|$, near the K points in the tight binding approximation. In the energy-momentum space close to the two Dirac points $\mathbf{K}$ and $\mathbf{K'}$, the retarded (advanced) bare Green's function [35] corresponding to the non-interacting Hamiltonian $H_0$ is a square matrix of order two in pseudo-spin space, given by

$$G_0^{A/R}(\epsilon(\mathbf{k}), \mathbf{p}) = \frac{1}{(\epsilon(\mathbf{k}) \mp i0) \mathbb{1} - v_f \sigma . \mathbf{p}} = \frac{\epsilon(\mathbf{k}) \mathbb{1} + v_f \sigma . \mathbf{p}}{(\epsilon(\mathbf{k}) \mp i0)^2 - v_f^2 p^2} \quad (2)$$

The figure 1(a) illustrates the Feynman diagram representing the electron propagator through equation (2). Similarly the other diagrams in Figure1 depict the following quantities discussed below, (b) Phonon propagator, (c) electron-impurity scattering, (d) e-ph self-energy, (e) Vertex correction of the e-ph interaction by impurity scattering.

The second term in the Hamiltonian is the non-interacting phonon Hamiltonian, $H_p = \sum_q \hbar\omega_q \, b_q^\dagger b_q$ where $\omega_q = v_{ph}|q|$ is linear phonon dispersion relation, $v_{ph}$ is longitudinal sound velocity, $b_q(b_q^\dagger)$ are phonon annihilation (creation) operators and $\mathbf{q}$ is the phonon momentum. The advanced (retarded) phonon bare Green's function $\Pi_0^{A/R}(\omega, q)$ shown in figure 1(b) is written as

$$\Pi_0^{A/R}(\omega, q) = \frac{2\omega_q}{\omega^2 - \omega_q^2 \mp i0} \qquad (3)$$

The third term in equation (1), $H_{dis} = \sum_{ri} \psi_i^\dagger(r) \, U(r) \psi_i(r)$, represents the total sum of randomly distributed scalar disorder potentials, $U(r) = \sum_j V(r - r_j)$. When the disorder $U(r)$ is smooth at the atomic scale it can be effectively treated as delta like scattering centers $U(r) = u_0 \delta(r - r_i)$ yielding the electron-impurity scattering time, $\tau \equiv l/v_f \propto v_g$, where $l$ is the mean free path and $v_g = \frac{4\epsilon_f}{2\pi\hbar^2 v_f^2}$ is the graphene density of states at the Fermi level, in which $\epsilon_f$ is the Fermi energy. The figure1c represents the Feynman diagram for a single electron-impurity scattering event. The impurity averaged advanced (retarded) Green's function [20] in the energy-momentum space corresponding to Hamiltonian $H_{dis}$ is simply obtained by substituting $\epsilon(k) \to \epsilon(k) \mp \frac{i}{2\tau}$, where the factor $\frac{i}{2\tau}$ arises from broadening of the energy levels;

$$G^{A/R}(\epsilon(k), p) = \frac{\epsilon(k) \mp \frac{i}{2\tau} + v_f \boldsymbol{\sigma}.\boldsymbol{p}}{\left(\epsilon(k) \mp \frac{i}{2\tau}\right)^2 - v_f^2 p^2} \qquad (4)$$

This disorder average broadening is the first effect of disorder arising from the e-ph interaction on the self-energy of electrons. The fourth term in the effective Hamiltonian accounts for the e-ph interaction arising from two sources the e-ph interaction due to scalar($V_S$) deformation potential, $H^1{}_{e-ph} = \sum_{k'kq} \psi_{k'}^\dagger V_S \psi_k (b_q + b_{-q}^\dagger)$ and that due to vector($V_V$) potential term, $H^2{}_{e-ph} = \sum_{k'kq} \psi_{k'}^\dagger V_V \psi_k (b_q + b_{-q}^\dagger)$. The $V_S$ is the DP coupling matrix element given by, $V_S = D^{DP}(\hat{e}.\overline{q})/(2\rho\omega_q \hbar^{-1})^{1/2}$, in which $D^{DP}$ is the scalar DP vertex of the e-ph interaction, $\hat{e}$ is the phonon polarization vector and $\rho$ is the graphene sheet mass density. The $H^2{}_{e-ph}$ contains $V_V = \left(v_f \sigma_i. \mathbf{A(r)}\right)$ where $\mathbf{A(r)}$ is the random vector potential that couples to velocity, which as mentioned earlier arises from a gauge field originating from the strain due to ripples. The vector DP coupling matrix element evaluates out as, $V_V = D^{VP} \sqrt{\frac{\hbar}{2\rho\omega_q}} \, iq \, e^{2i\varphi_q}$, where $D^{VP}$ is the vector DP vertex of the e-ph interaction equal to $D^{VP} = \frac{v_f \beta\hbar}{2a}$, and has a constant value.

The Drude DC resistivity in graphene due to electron-acoustic phonon interaction is given by [21],

$$\rho_g = \frac{(\frac{v_f^2}{2} e^2 v_g)^{-1}}{\tau_{e-ph}} \qquad (5)$$

in which, $\tau_{e-ph}$ is e-ph relaxation time. The Keyldysh Green's function which is versatile enough to capture the interface between e-ph and electron-impurity scatterings, i.e. the vertex correction of the e-ph vertices by disorder [31] is used here for calculating the e-ph relaxation rate. The full electron Keyldysh Green's function $\check{G}_0(\epsilon(\boldsymbol{k}), \boldsymbol{p})$ for equation (1) is

$$\check{G}_0(\epsilon(\boldsymbol{k}), \boldsymbol{p}) = \left(\check{G}_0(\epsilon(\boldsymbol{k}), \boldsymbol{p}) - \check{\Sigma}(\epsilon(\boldsymbol{k}), \boldsymbol{p})\right)^{-1} \qquad (6)$$

Where $\check{\Sigma}$ is the electronic Keyldysh self-energy, and $\check{G}_0$ is the bare matrix Green's function in Keyldysh space [31];

$$\check{G}_0(\epsilon(\boldsymbol{k}), \boldsymbol{p}) = \begin{pmatrix} G_0^R(\epsilon(\boldsymbol{k}), \boldsymbol{p}) & G_0^K(\epsilon(\boldsymbol{k}), \boldsymbol{p}) \\ 0 & G_0^A(\epsilon(\boldsymbol{k}), \boldsymbol{p}) \end{pmatrix} \qquad (7)$$

In which,

$$G_0^K(\epsilon(\boldsymbol{k}), \boldsymbol{p}) = [1 - 2n(\epsilon(\boldsymbol{k}), T_e)][G_0^R(\epsilon(\boldsymbol{k}), \boldsymbol{p}) - G_0^A(\epsilon(\boldsymbol{k}), \boldsymbol{p}) \qquad (8)$$

The $n(\epsilon(\boldsymbol{k}), T_e)$ is the electronic Fermi-Dirac distribution function at $T = T_e$. Similarly the full phonon Keyldysh Green's function $\Pi^K(\omega, \boldsymbol{q})$ is

$$\Pi^K(\omega, \boldsymbol{q}) = [1 + 2N(\omega, T_{ph})][\Pi^R(\omega, \boldsymbol{q}) - \Pi^A(\omega, \boldsymbol{q}) \qquad (9)$$

Where,

$$\Pi_0(\omega, \boldsymbol{q}) = \begin{pmatrix} \Pi_0^R(\omega, \boldsymbol{q}) & \Pi_0^K(\omega, \boldsymbol{q}) \\ 0 & \Pi_0^A(\omega, \boldsymbol{q}) \end{pmatrix} \qquad (10)$$

and $N(\omega, T_{ph})$ is the phononic Bose-Einstein distribution function at $T = T_{ph}$. The relaxation rate due to e-ph interaction is calculated as a variation of collision integral, $I_{e-ph}(p, \epsilon(\boldsymbol{k}))$;

$$\frac{1}{\tau_{e-ph}} = -\frac{\delta I_{e-ph}(p, \epsilon(\boldsymbol{k}))}{\delta n(\epsilon(\boldsymbol{k}))} \quad (11)$$

The collision integral tells the rate of change of electronic phase-space of distribution function $[n(p, \epsilon(\boldsymbol{k}); t)]$ due to the emission and absorption of acoustical phonons, and is given by the same expression as that for diffusive metals[13], except for the inclusion of matrix structure of e-ph and electron-electron (el-el) vertices in the pseudo spin space for graphene,

$$I_{e-ph}(p, \epsilon(\boldsymbol{k})) = \left[\frac{dn(p, \epsilon(\boldsymbol{k}))}{dt}\right] = -4\tau \int \frac{d\omega dq}{(2\pi)^3} \frac{\mu^2}{|\epsilon(q)|^2} Im\Pi_0^{A/R}(\omega, \boldsymbol{q}) \eta(\omega, \epsilon(\boldsymbol{k})) Re\left\{\frac{\check{h}_n(q, \omega)}{1 - \check{h}_n(q, \omega)}\right\} (12)$$

Where $\mu = D^{DP}(D^{VP})$, $\varepsilon(q)$ is the static dielectric function, $Im\Pi_0^{A/R}(\omega, \boldsymbol{q})$ is the imaginary part of retarded phonon Green's function given by equation (3), $\eta(\omega, \epsilon(\boldsymbol{k})) = N(\omega, T_{ph})n(\epsilon(\boldsymbol{k}), T_e)\{1 - n(\epsilon(\boldsymbol{k}) + \omega, T_e)\} - \{1 + N(\omega, T_{ph})\}\{1 - n(\epsilon(\boldsymbol{k}), T_e)\}n(\epsilon(\boldsymbol{k}) + \omega, T_e)$ is the combination of electronic and phononic distribution functions appropriate for phonon emission and absorption processes. The factor $Re\left\{\frac{\check{h}_n(q, \omega)}{1 - \check{h}_n(q, \omega)}\right\}$ is the real part of the response function which accounts for the effect of disorder through the vertex correction of the e-ph vertex due to impurity scattering, as shown in figure1(e). The function $\check{h}_n(q, \omega)$ is an integral over impurity averaged electron Green function in two dimensions [13],

$$\check{h}_n(\boldsymbol{q}, \omega) = \frac{g}{2\pi\nu_g\tau} \int \frac{d^2p}{(2\pi)^2} G^A(\boldsymbol{p}, \epsilon(\boldsymbol{k})) G^R(\boldsymbol{p} + \boldsymbol{q}, \epsilon(\boldsymbol{k}) + \omega) (13)$$

Where $G^A(\epsilon(\boldsymbol{k}), p) = [G^R(\epsilon(\boldsymbol{k}), p)]^*$. In case of graphene, modifications in equation (13) arise because of the pseudo spin space matrix structure which has been reported in detail [20]. As mentioned before we work in the weakly disordered regime $k_F l >> 1$, in which the characteristic temperature limit below which disorder effects become important is defined as $T_{dr} = \hbar v_{ph}/k_B l$, where $v_{ph}$ is the acoustic phonon velocity in graphene. This temperature limit $T_{dr}$ below which disorder effects dominate is necessarily well within the Bloch-Gruneisen temperature limit, $T_{BG} = 2\hbar v_{ph} k_F/k_B$. The evaluation of equation (12) after integrating over the phonon frequency and then after averaging over the electron momentum in the temperature range, $\frac{v_{ph}}{v_f} T_{dr} < T \ll T_{BG}$, the corresponding collision integral for the DP and VP coupling respectively have been reported [20], as under,

$$I_{e-ph}^{DP}(\epsilon) = -\tau \int \frac{dq}{(2\pi)^2} \eta(\omega_q, \epsilon) \mu^2 \left[ \frac{1}{q^2 l^2}\left(\frac{1}{\sqrt{1+q^2 l^2}} - 1\right) + \left(\frac{1}{q^2 l^2} + \frac{1}{\sqrt{1+q^2 l^2}}\right) \right] (14)$$

and

$$I_{e-ph}^{VP}(\epsilon) = -\tau \int \frac{dq}{(2\pi)^2} \eta(\omega_q, \epsilon) \mu^2 \left[ \left(\frac{1}{2(1+q^2 l^2)} + \frac{1}{\sqrt{1+q^2 l^2}}\right) - \frac{1}{2}\left(\frac{1}{q^2 l^2} - \frac{1}{q^2 l^2 \sqrt{1+q^2 l^2}}\right) \right] \quad (15)$$

In Ref. [20], the heat flux $P(T_e, T_{ph}) = v_g \int d\epsilon \, \epsilon_{e-ph}(p, \epsilon) = F(T_e) - F(T_{ph})$ has been calculated from the collision integral, where the functions $F(T_e)$ and $F(T_{ph})$ are energy control functions for electrons and phonons respectively. The $F(T_{ph})$ is given by the equation,

$$F(T_{ph}) = 4\tau v_g \int \frac{dq}{(2\pi)^3} \frac{\mu^2}{|\varepsilon(q)|^2} \text{Re}\left\{\frac{\ddot{\Pi}_n(q,\omega)}{1-\ddot{\Pi}_n(q,\omega)}\right\} [N(\omega_q, T_{ph})] (16)$$

After evaluating all these functions and substituting in equation (13), as shown in Ref.[20], one obtains the energy control function for DP coupling in the temperature range $\frac{v_{ph}}{v_f} T_{dis} < T \ll T_{BG}$, which we quote below for comparison with the expression for $\frac{1}{\tau_{e-ph}}$,

$$F^{DP}(T_{ph}) = 4\tau v_g \int_0^\infty dq \, q^3 \left[ \frac{1}{q^2 l^2}\left(\frac{1}{\sqrt{1+q^2 l^2}} - 1\right) + \left(\frac{1}{q^2 l^2} + \frac{1}{\sqrt{1+q^2 l^2}}\right) \right] [N(\omega_q, T_{ph})] \quad (17)$$

The e-ph relaxation rate at equilibrium at the Fermi surface from equation (10) is given by [13],

$$\frac{1}{\tau_{e-ph}} = -4\tau \int \frac{dq}{(2\pi)^3} \frac{\mu^2}{|\varepsilon(q)|^2} \text{Re}\left\{\frac{\ddot{\Pi}_n(q,\omega)}{1-\ddot{\Pi}_n(q,\omega)}\right\} [N(\omega_q, T_{ph}) + n(\epsilon(\boldsymbol{k}), T_e)] (18)$$

In the same approximation as that obtained for $F^{DP}$ and $F^{VP}$, we obtain the following pair of equations for $\frac{1}{\tau_{e-ph}^{DP}}$ and $\frac{1}{\tau_{e-ph}^{VP}}$,

$$\frac{1}{\tau_{e-ph}^{DP}} = \tau \int \frac{dq}{(2\pi)^2} \mu^2 \left[ \frac{1}{q^2 l^2}\left(\frac{1}{\sqrt{1+q^2 l^2}} - 1\right) + \left(\frac{1}{q^2 l^2} + \frac{1}{\sqrt{1+q^2 l^2}}\right) \right] [N(\omega_q, T_{ph}) + n(\epsilon(\boldsymbol{k}), T_e)] \quad (19)$$

$$\frac{1}{\tau_{e-ph}^{VP}} = \tau \int \frac{dq}{(2\pi)^2} \mu^2 \left[ \left(\frac{1}{2(1+q^2 l^2)} + \frac{1}{\sqrt{1+q^2 l^2}}\right) - \frac{1}{2}\left(\frac{1}{q^2 l^2} - \frac{1}{q^2 l^2 \sqrt{1+q^2 l^2}}\right) \right] [N(\omega_q, T_{ph}) + n(\epsilon(\boldsymbol{k}), T_e)]$$

$$(20)$$

Since we also include screening in the estimation of the e-ph relaxation rate, we briefly discuss here the screening dielectric functions in which we have made the calculations. We consider only the screening of DP coupling at long wavelengths and do not include VP screening as it does not induce any net charge density [20]. The screening of DP is defined in term of the static dielectric function $\varepsilon(q) = 1 + V_0(q)P(q)$, where $V_0(q) = 2\pi e^2/\varepsilon_0 q$ is the two-dimensional Fourier transform of bare Coulomb potential, $e$ is the electronic charge, $\varepsilon_0$ is dielectric constant and $P(q)$ is the polarization operator for graphene [39]. The Thomas-Fermi (TF) and the Random Phase Approximation (RPA) dielectric functions are generally used to determine the effect of electric field screening by electrons in solids. The TF model is valid for a small wave vector q>$k_F$ only. This shortcoming is overcome in RPA which reduces to TF in the small wave vector limit. The TF dielectric function is given by $\varepsilon(q) = 1 + \frac{2\pi e^2}{\varepsilon_0 q} v_g$. The dielectric function in the RPA can be calculated using the polarization function for the graphene which is given by [37],

$$P(q) = v_g \left[ 1 - \theta(q - 2k_F)\left( \frac{q}{4k_F} Sin^{-1}\left(\frac{2k_F}{q}\right) + \frac{1}{2}\sqrt{1 - \left(\frac{2k_F}{q}\right)^2}\right)\right] (21)$$

### 3. Analytical and numerical results and discussion

At low temperatures the wavelength of thermal phonon is comparable to or longer than the electronic mean free path and therefore the electronic disorder can strongly modify the e-ph coupling. The temperature limit $T_{dr}$ below which disorder effects dominate comes out to be ($T_{dr} \sim$ 7.6 K, 1.5 K, 0.3K) for mean free paths of respectively, (20 nm, 100 nm, 0.5µm), with $v_{ph}$=2×10⁶ cm/S. The carrier density corresponds to Bloch-Gruneisen temperature limit, $T_{BG}$. Our main results for the relaxation rates from equation (18) for graphene in the temperature range $\frac{v_{ph}}{v_f}T_{dr} < T \ll T_{BG}$ are evaluated as follows. The equations (19) & (20) are quite intractable analytically even in the absence of screening but are amenable to analytic evaluation in the extreme diffusive limit ($l \to 0$) and clean limit ($l \to \infty$) yielding results in closed form. The closed form result for the case of DP coupling from equation(19) in the dirty and clean limit is represented, respectively by equations(22a) and (22b), as under;

$$\frac{1}{\tau_{e-ph}^{DP}} = \frac{2 D^{DP^2} k_B T}{\pi \rho v_f v_{ph}^2 \hbar^2 l} (22a)$$

$$\frac{1}{\tau_{e-ph}^{DP}} = \frac{\pi D^{DP^2} k_B^2 T^2}{4 \rho v_f v_{ph}^3 \hbar^3} (22b)$$

Similarly the closed form solution for the case of VP coupling from equation (20) in the dirty and clean limit is represented, respectively by equations (23a) and (23b), as given below;

$$\frac{1}{\tau_{e-ph}^{VP}} = \frac{7 D^{VP^2} l \zeta(3) k_B^3 T^3}{2 \pi \rho_f v_{ph}^4 \hbar^4} (23a)$$

$$\frac{1}{\tau_{e-ph}^{VP}} = \frac{\pi D^{VP^2} k_B^2 T^2}{4 \rho v_f v_{ph}^3 \hbar^3} (23b)$$

We notice that in all the four above derived analytical results the temperature exponent is reduced as compared to the $T^4$ result obtained in BTF by either one, two or three counts. The equation (22a) shows a linear relaxation rate with temperature with DP e-ph coupling and is in good agreement at high temperatures with the earlier reported study for the disordered graphene [6]. The $T^2$ temperature dependency of the DP scattering rate in the clean limit in equation (22b) obtained through the Keldysh method is in accordance with the reported finding that the experimental crossover of the resistivity curve to the high temperature linear behavior appears to be closer to a $T^2$ behavior rather than the $T^4$ result obtained in BTF [21, 36]. This result is in contrast to the result for heat flux obtained in the Keldysh formalism reported in Ref. [20], where for the DP coupling the exponent is reduced to $T^3$behavior while that of VP coupling it is enhanced to $T^5$ behavior. In other words the heat flux is enhanced in case of DP while it is suppressed in case of VP for $T<T_{dr}$. Comparing equations(20b) and (21b) for DP and VP coupling respectively in the pure limit ($T>T_{dr}$) with the BTF pure limit result ($T^4$), we observe that in both the cases the temperature power exponent is reduced by the same count ($T^2$), implying that the relaxation rates are enhanced by the same power exponent. This result is also in contrast to the result reported for heat flux where for the DP and VP coupling no change occurs in the power exponent, and it shows the same BTF $T^4$ dependence [20]. However since the VP coupling constant is one order of magnitude smaller than the DP constant the scattering rate due to VP coupling will be much lesser that that of DP coupling relaxation rate. If we include screening in the TF approximation for the case of DP coupling then the power exponent increases by $T^2$ in both the limits. The vector potential is not screened as it does not induce any net charge density [20, 32]. All the obtained resistivity analytical results have been tabulated in Table 1 along with the reported heat flux analytical results in the same limit in the Keldysh technique for the ease of comparison [20].

However the approximate analytical equations (22) and (23) cannot be a replacement for a full numerical evaluation of the concerned equations. In order to see the complete effects of disorder on DP as well as VP coupling as a function of temperature, mean free path and carrier density we took recourse to numerical computation.

| | VP coupling | | DP coupling | | | |
|---|---|---|---|---|---|---|
| | Diffusive limit ($l\rightarrow0$) | Clean limit ($l\rightarrow\infty$) | Diffusive limit ($l\rightarrow0$) | | Clean limit ($l\rightarrow\infty$) | |
| | | | Unscreened | Screened ($\varepsilon_0^2/\alpha^2$) | Unscreened | Screened ($\varepsilon_0^2/\alpha^2$) |
| $\rho_g$ | $\dfrac{7C_1 l\zeta(3)k_B^3 T^3}{2k_f e^2 v_{ph}^4 \hbar^3}$ | $\dfrac{\pi^2 C_1 k_B^2 T^2}{4k_f e^2 v_{ph}^3 \hbar^2}$ | $\dfrac{2C_2 k_B T}{4k_f e^2 v_{ph}^2 \hbar l}$ | $\dfrac{7C_2 \zeta(3)k_B^3 T^3}{8k_f^3 e^2 v_{ph}^4 \hbar^3 l}$ | $\dfrac{\pi^2 C_2 k_B^2 T^2}{4k_f e^2 v_{ph}^3 \hbar^2}$ | $\dfrac{\pi^4 C_2 k_B^4 T^4}{32k_f^3 e^2 v_{ph}^5 \hbar^4}$ |
| $F^{\#}$ | $\dfrac{30C_1 k_f l\zeta(5)k_B^5 T^5}{\pi^2 v_{ph}^4 \hbar^5}$ | $\dfrac{\pi^2 C_1 k_f k_B^4 T^4}{v_{ph}^3 \hbar^4}$ | $\dfrac{2C_2 \zeta(3)k_f k_B^3 T^3}{v_{ph}^2 \hbar^3 l}$ | $\dfrac{6\pi^2 C_2 \zeta(5)k_B^5 T^5}{\pi^2 v_{ph}^4 \hbar^5 l}$ | $\dfrac{\pi^2 C_2 k_f k_B^4 T^4}{v_{ph}^3 \hbar^4}$ | $\dfrac{2\pi^4 C_2 k_B^6 T^6}{63 v_{ph}^5 \hbar^6}$ |

#Ref. [20]. Chen W and Clerk A A 2012 *Phys. Rev. B***86** 125443

**Table 1**.The temperature($T$) , mean free path($l$) and carrier concentration($n$) dependencies of resistivity $\rho_g(T, l, n)$ and  energy loss rate $F\ (T, l)$ are summarized above for electron–acoustic phonon interaction via Vector potential coupling and Deformation-potential coupling (unscreened and TF screened) in graphene. In which $\varepsilon_0$ is dielectric constant,$\alpha = \dfrac{e^2}{v_f \hbar}$ is coupling constant, and $C_1 = \dfrac{D^{VP^2}}{\rho v_f^2}$, $C_2 = \dfrac{D^{DP^2}}{\rho v_f^2}$.

The resistivity of graphene has been computed as a function of temperature, mean free path and carrier density for disordered graphene by using the following material parameter; $D^{DP}$=19 eV,  $D^{VP}$=1.5eV,  $v_f$=1×10$^8$cm/sec, electron density $n$=2×10$^{12}$cm$^{-2}$,$v_{ph}$=2×10$^6$cm/sec, $\rho$=7.6×10$^{-8}$g/cm$^2$, $\varepsilon_0$=5.7 effective dielectric constant for graphene on SiO$_2$substrate [24,37]. The thermal decoupling of electrons from the crystal lattice in most materials takes place at temperatures of order of a few Kelvin [33, 38]. The comparison with the approximate analytical results is most appropriate in the sub-Kelvin temperatures since the approximate formulae for collision integral, equations (14) and (15), have been derived in the strong impurity limit, where $ql<<1$ or $T<T_{dr}$.

In figures 2-4, we plot the numerically and analytically calculated graphene resistivity from equation (5), substituting the relaxation time from equations (19) & (20) and equations (22) & (23) for DP and VP couplings, respectively, as functions of temperature, mean free path and carrier density. To figure out the accuracy of obtained approximate analytical results with the complete numerical evaluation of the concerned equation, we in figure 2(a) plot the numerically and analytically computed resistivity due to unscreened DP el-ph coupling from analytical equations (22a)-(22b) and numerical equation (19), respectively. We observe from this plot that the DP numerical curve-C lies almost within the two analytical curves-A & J. The analytical dirty limit Curve-A of equation (22a) and numerical Curve-C from equation (19) lie close together below $T<T_{dr}$ but depart at higher temperatures, and that the numerical curve-C approaches the analytical clean limit DP curve-J of equation (22b) at higher temperatures. In rest of the plots we do not include analytical results of either the DP or VP couplings in the clean limit and make a comparison with the numerical results only with the analytical result in the dirty limit (curves–A & B). The figure 2(b) shows the plotted numerically calculated resistivity versus temperature and its comparison with the analytical result in dirty limit due to unscreened DP and VP coupling as a function of the temperature with different mean free path lengths ($l$= 20nm, 100nm and .5μm) at a fixed $n$=2×10¹² cm⁻². The following conclusions can be drawn from the juxtaposed numerically computed unscreened DP and VP coupling curves in figure 2(b): (i) the analytical dirty limit VP coupling (curve-B) is suppressed by disorder while the DP coupling (curve-A) is enhanced, as it is described by a reduced power law (linear $T$ dependence for unscreened DP) and thus the relaxation rate dominates the VP at low temperatures ($T^3$ dependence), (ii) At lower temperatures, the temperature exponent of resistivity due to both the couplings (DP &VP) vary with mean free path. For $l$=20nm, the resistivity for DP coupling varies as $T$ and for VP coupling varies as $T^3$ at low temperatures, but both tend to $T^2$ behavior at higher temperatures. With increasing mean free path a contrasting behavior in resistivity from both the couplings is noticed, as for $l$= .5μm resistivity for DP coupling shows an increase in exponent (from $T$ to $T^{1.5}$) but for VP coupling the exponent declines from ($T^3$ to $T^{2.5}$) with a cross over temperature that shifts to lower temperature values with increasing disorder, as can be observed from figure 2(b). To investigate the behavior of electronic screening, we plotted in figure 2(c) the variation of resistivity with temperature for different values of mean free path ($l$= 20nm, 100nm and .5μm) at fixed value of $n$=2×10¹² cm⁻². The inclusion of screening in the form of TF & RPA dielectric functions affect the resistivity more at lower temperatures ($T<2$K) where it goes as $T^3$ but for higher temperatures the TF screening behavior approaches the $T^2$ behavior of unscreened DP, however the RPA screening shows a non-monotonic behavior at higher temperatures. The figure 2(d) is a comparative plot of screened and unscreened resistivity with temperature for different values of mean free path ($l$= 20nm, 100nm and .5μm) at fixed value of $n$=2×10¹² cm⁻². Here we can easily see from the plots that the same power $T^3$ dependence at low temperature for all the curves-D, E and F except the unscreened DP curve-C which exhibits a linear $T$ behavior of equation (20a) at lower temperatures. Also it can be noticed from figure 2d that (i) both the DP and VP couplings at low temperatures make nearly equal contributions to resistivity when both disorder and screening are included, (ii) the TF and RPA screening contribute equally well below the BG temperature limit less than $T<20$ K but depart from each after at higher temperatures with the TF screening curve-D approaching the unscreened DP curve-C and the RPA screening curve approaching the VP curve-E. This means that the TF dielectric function →1 and that the TF screening is not effective at higher temperatures.

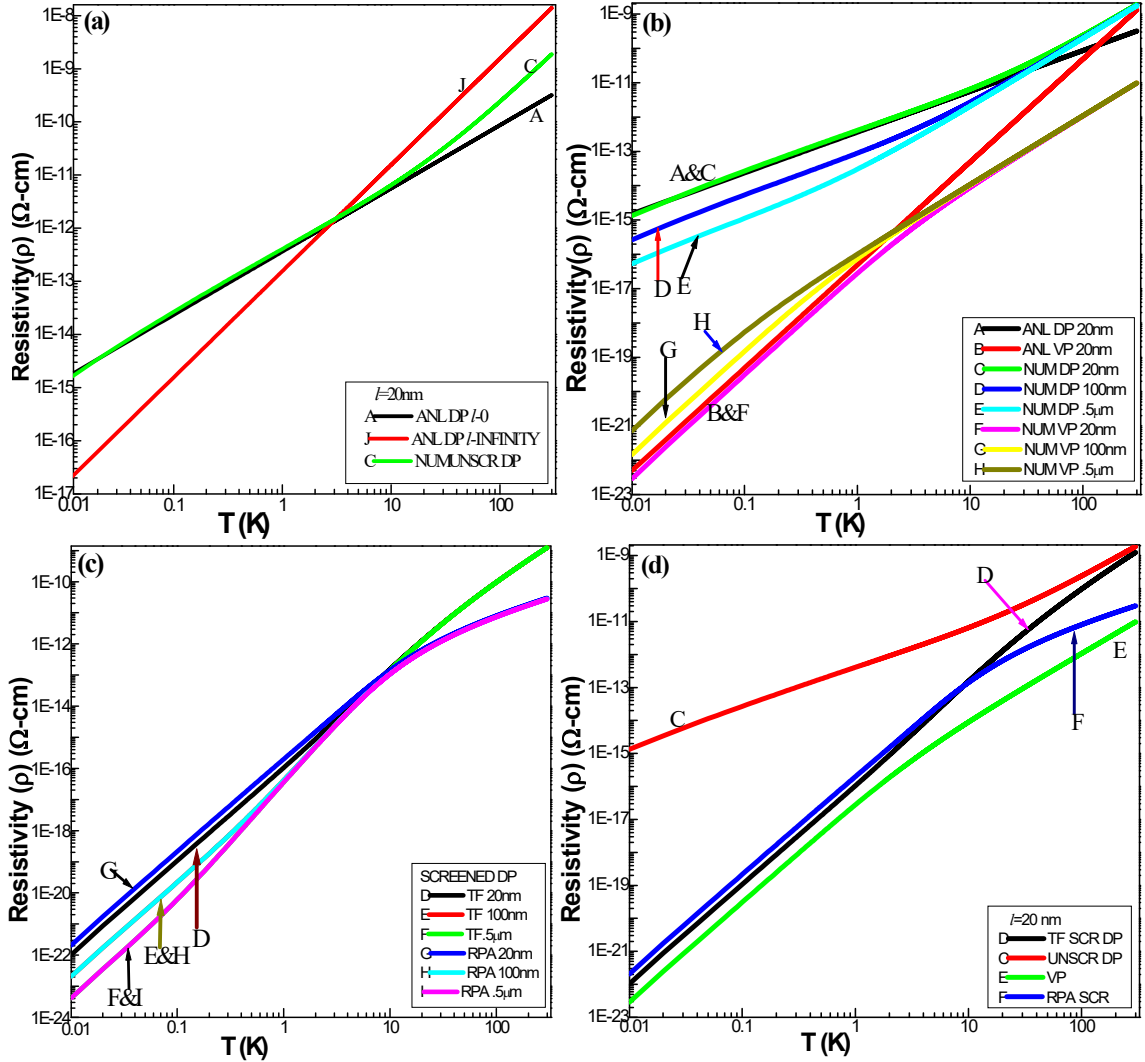

**Figure 2**. Disordered graphene resistivity due to e-ph coupling as a function of temperature at a fixed $n=2\times10^{12}$ cm$^{-2}$. The labelling of curves is as follows: Curves-A, B represent analytical (ANL), unscreened (UNSCR) DP and VP in dirty limit from equations (22a), (23a), respectively, and & curve-J is ANL UNSCR DP in clean limit from equation (22b). Figure (a) shows a comparative plot of approximate analytical results for unscreened DP coupling, along with the numerical resistivity at $l=$ 20nm, (b) exhibits a comparative plot with different mean free path lengths for unscreened DP and VP coupling, Curves-C, D & E depict numerical unscreened DP resistivity from equation (19), Curves-F, G & H exhibit numerically computed unscreened VP resistivity from equation (20), (c) depicts the screened DP behavior under TF and RPA where curves-D, E & F represent TF screening while curves-G, H & I represent RPA screening,, (d) compares the screened and unscreened resistivity in DP and VP coupling, for $l=$ 20nm where DP curves-C, D & F represent Unscreened, TF screened, and RPA screened, respectively while curve-E is VP unscreened numerical curve.

.

To elicit the explicit role of disorder or elastic mean free path $l$ on the resistivity we plotted in figures 3(a)-3(d) the variation of resistivity with mean free path for different values of temperature (i.e., $T=.1K, T=1K$ and $T=10K$) at a fixed value of $n=2\times10^{12}$ cm$^{-2}$. As done figure 2(a)-2(d) the numerical curves have been computed using equations (19) and (20) and the analytical curves from that given by equations (22a) & (23a), in the strong impurity limit. These four figures 3(a)-3(d) validate the point drawn from the previous figures 2(a)-2(d), that at very low temperatures with DP coupling (both for screened and unscreened & for analytical and numerical expressions) we find that the resistivity decreases with increase in mean free path and that it varies inversely with $l$. In contrast to this the resistivity from VP coupling increases directly with increasing mean free path or decreasing amount of disorder. In case of unscreened DP coupling from figure 3(a) we find no change in mean free path exponent for all the plotted curves-A to E in the whole covered temperature range ($T<T_{dr}$ and $T>T_{dr}$). However for VP coupling the resistivity curves in figure 3b vary from $l^1$ to $l^{0.7}$ and $l^1$ to $l^{0.05}$ for $T<T_{dr}$ and $T>T_{dr}$, respectively. The figure 3(c) is a plot for screened behavior of DP coupling. We observe that the TF and RPA screened DP coupling curves lie in close proximity in the whole plotted range, which means the TF approximation captures very well the effect of static screening. The numerical curves for TF and RPA electronic screened DP coupling varies in plotted range as $l^{-1}$ to $l^{-0.7}$ and $l^{-0.94}$ to $l^{-0.01}$ for $T<T_{dr}$ and $T>T_{dr}$, respectively. This implies that screening has greater influence at higher temperatures ($T>T_{dr}$) and mean free path (that is lesser disorder) where it results in the reduction of scattering rate and thereby the resistivity in disorder graphene. The differences between all three, that is DP unscreened, DP and VP screened, are best highlighted in the figure 3(d).

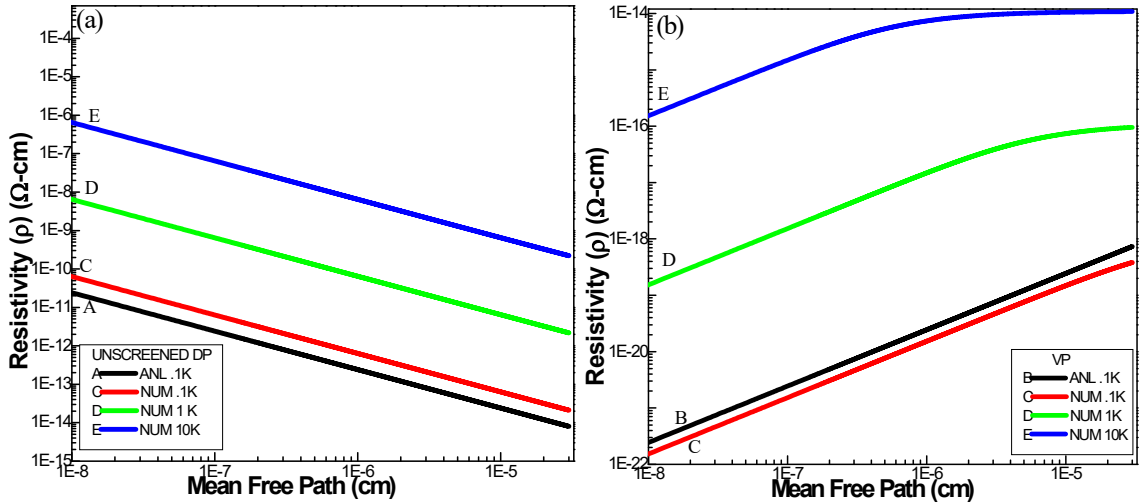

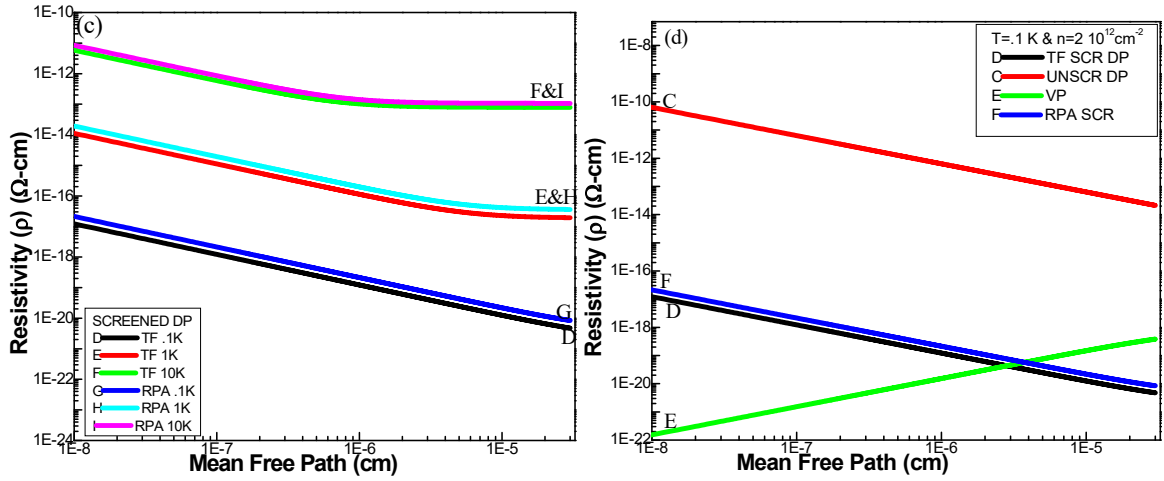

**Figure 3.** Disordered graphene resistivity as a function of mean free path with different temperature and fixed $n=2\times10^{12}$ cm$^{-2}$. (a) plot of analytical and numerical unscreened DP resistivity at different temperatures ($T=$ .1K, 1K and 10K) where curve-A represents analytical unscreened DP from equation (22a) while numerical curves-C, D & E are from equation (19), (b) plot of analytical and numerical unscreened VP resistivity at different temperatures, where curve-B represents analytical unscreened VP from equation (23a) while numerical curves-C, D & E are from equation (20), (c) plot of screened DP resistivity under TF and RPA where curves-D, E & F represent TF screening while curves-G, H & I represent RPA screening, (d) comparative plot of the screened and unscreened resistivity in DP and VP coupling, at $T=0.1$K where DP curves-C, D & F represent Unscreened, TF screened, and RPA screened, respectively while curve-E is VP unscreened numerical curve.

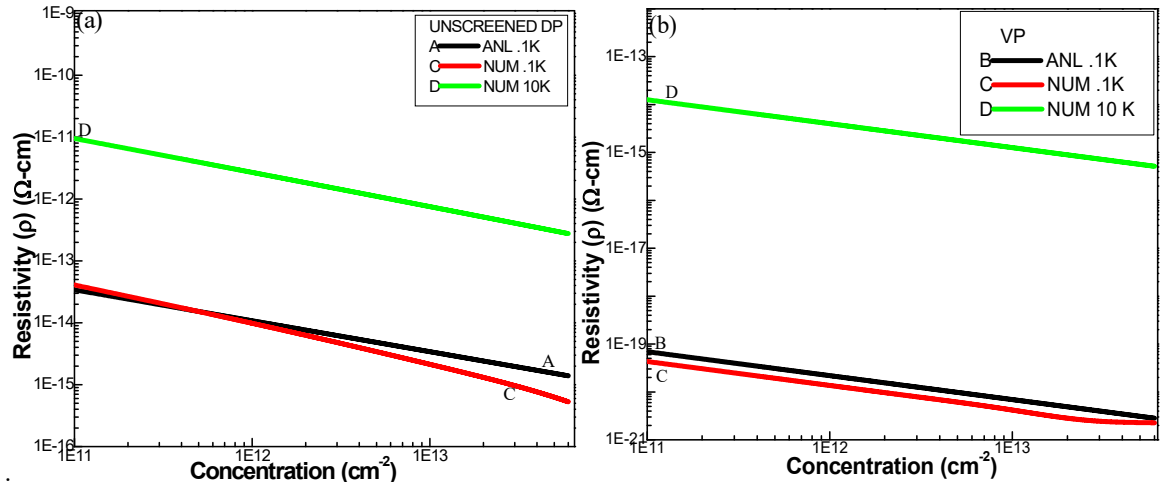

.

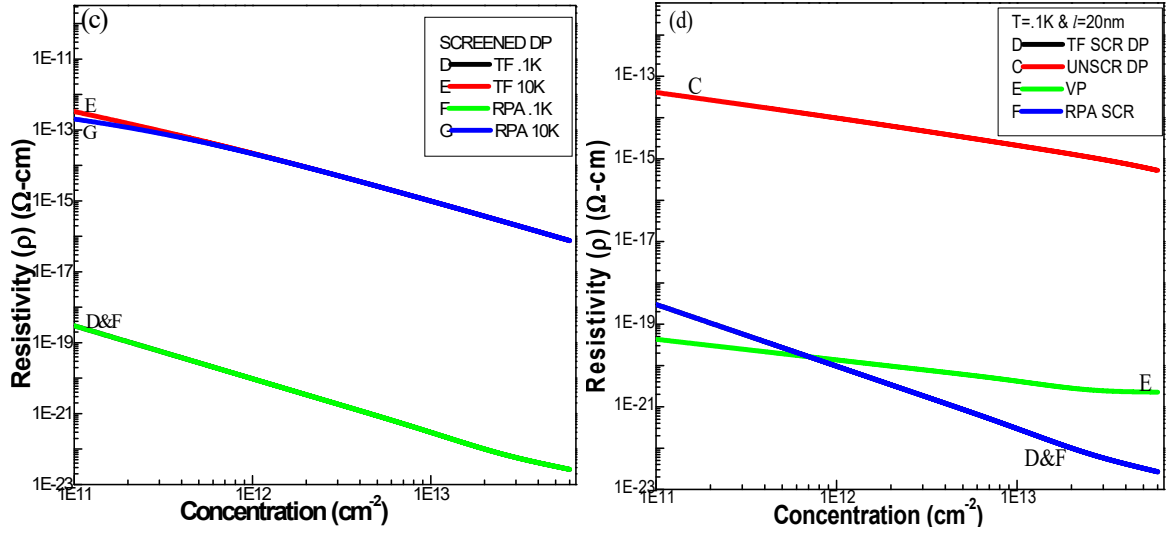

**Figure 4.** Disordered graphene resistivity as a function of concentration with different temperatures at a fixed $l$=20nm. (a) plot of analytical and numerical unscreened DP resistivity at different temperatures ($T$= .1K and 10K) where curve-A represents analytical screened DP and curves-C & D are numerical DP screened, (b) plot of analytical and numerical unscreened VP resistivity at two different temperatures where curve-B represents analytical unscreened VP and curves-C & D are numerical unscreened VP curves, (c) plot of screened DP resistivity under TF and RPA at two different temperatures where curves-D & E are TF screened DP curves and curves-F & G are RPA screened DP curves (d) comparative plot of the screened and unscreened resistivity in DP and VP coupling, at $T$=0.1K where DP curves-C, D & F represent Unscreened, TF screened, and RPA screening, respectively while curve-E is VP unscreened numerical curve.

The variation of resistivity with carrier concentration (at $T$ = .1K, $T$=10K & $l$=20nm) is depicted in figures 4(a) to 4(d). The figures 4(a) & 4(b) plotted respectively for DP and VP coupling, illustrate the behavior of analytical and numerically computed curves using equations (22) & (23) and equations (19) &(20) for DP and VP couplings. Similar to the figure 3(a) behavior of variation with increasing mean free path here also the same trend that the resistivity declines with decreasing carrier density is noticed, albeit with a different power dependency. The power dependence on carrier density of the unscreened DP coupling numerical curve-C for $T$<$T_{dr}$ and curve-D for $T$>$T_{dr}$ varies as $n^{-0.5}$to $n^{-0.1}$ in the plotted range while the numerical unscreened VP coupling curve-C for $T$<$T_{dr}$ and curve-D for $T$>$T_{dr}$ varies as $n^{-0.5}$in plotted range. Most of the experimental and theoretical work in graphene, carried out in diffusive limit report the conductivity (σ) to exhibit a linear to sub-linear to $n$- independent behavior [32-33, 40]. Studies also exist in which $\sigma = nlnn^2$dependence have been reported. In case of ballistic transport $\sigma = \sqrt{n}$has been reported [33]. From figure 4c we observe that the TF and RPA screened DP coupling curves merge for the plotted range when $T$=.1K<$T_{dr}$ but begin to part at lower carrier density at higher temperatures that is when $T$=10K>$T_{dr}$. This means that the TF dielectric function is as good as the more comprehensive RPA function at higher carrier densities. The numerical curves for both the TF and RPA DP coupling varies in plotted range as $n^{-1.5}$ to $n^{-1}$ and $n^{-1.1}$ to $n^{-1.5}$ for $T$<$T_{dr}$ and $T$>$T_{dr}$, respectively.

So, we can say from this theoretical investigation using the Keyldysh non equilibrium green's function method for the scattering rate in the diffusive limit of disordered graphene as a function of temperature, mean free path, concentration and screening, that the presence of disorder significantly modifies the magnitude and behavior of e-ph scattering and thereby the resistivity behavior. We believe that the results obtained will boost further investigations to advance the understanding of transport phenomena in disordered graphene.

## 4. Conclusion

We have investigated the temperature, disorder, and concentration dependence of resistivity near the Dirac point of the disordered graphene in the Keldysh non-equilibrium Green's function approach for the cases of the Deformation and Vector electron-phonon coupling. From this formalism, we could reproduce the analytical results obtained in the asymptotic limit of clean ($T^2$) and strong impurity limit (T) for unscreened DP coupling. We obtain different analytical power dependencies in these limits in the case of VP coupling. The full numerical evaluation of the expression modifies these results and shows that the temperature dependence of resistivity is significantly affected by the presence of disorder. Hence it can be concluded that the electronic disorder influences the electron-phonon interaction in both DP and VP coupling mode, and affects the temperature, mean free path and concentration dependence of resistivity. This is in effect a bearing of the chiral nature of free carriers in graphene and its influence on the diffusive dynamics of the charge carriers.